THE EUROPEAN
PHYSICAL JOURNAL PLUS

Review

# The SPHERE view of multiple star formation

Raffaele Gratton[1,a] , Silvano Desidera[1,b], Francesco Marzari[2,c], Mariangela Bonavita[3,d]

[1] Osservatorio Astronomico di Padova, INAF, Vicolo dell'Osservatorio 5, 35122 Padova, Italy
[2] Dipartimento di Fisica e Astronomia Galileo Galilei, Università di Padova, Via Marzolo 8, 35121 Padova, Italy
[3] School of Physical Sciences, The Open University, Walton Hall, Milton Keynes MK7 6AA, UK

**Abstract** While a large fraction of the stars are in multiple systems, our understanding of the processes leading to the formation of these systems is still inadequate. Given the large theoretical uncertainties, observation plays a basic role. Here we discuss the contribution of high contrast imaging, and more specifically of the SPHERE instrument at the ESO Very Large Telescope, in this area. SPHERE nicely complements other instruments such as Gaia or ALMA—in detecting and characterizing systems near the peak of the binary distribution with separation and allows to capture snapshots of binary formation within disks that are invaluable for the understanding of disk fragmentation.

## 1 Introduction

More than 70% of massive early-type stars [1, 2], 50–60% of solar-type stars [3–6]), and 30–40% of the M-stars [7–9] are in multiple systems. An even higher fraction of binaries has been observed in low-density star forming regions [10–12], though it is yet unclear if this excess extends over all masses or separations, or is rather limited to the smallest masses or the widest separation range. Given the large frequency of multiple systems, understanding how they form is a basic step in our comprehension of the star formation process.

Leaving aside the case of stars with very large masses [13], various mechanisms likely concur in the formation of multiple stars of small and intermediate masses (see, e.g., [5, 14, 15]). Most likely, binaries with separations larger than 500 au mainly form through (turbulent) core fragmentation of clouds [16–18] and those having separation shorter than 500 au mainly by disk fragmentation [19], with the two mechanisms not to be thought as mutually exclusive. L1448 IRS3 is an on-sky example of how both mechanisms may likely work simultaneously on different scales [20]. However, the value of the separation where either of the two processes prevail is not well determined and likely depends on the mass of the stars [21, 22] and on the environment (we focus here on low-density regions). Disk fragmentation depends on the disk-to-star mass ratio, that is expected to be larger during early phases of formation and in more massive objects [21, 23, 24]. Interestingly, within disks mass accretion onto the secondary may be favored with respect to accretion onto the primary [25, 26], at least so far the binary orbit is coplanar to the disk. This may produce equal mass binaries if the disk survives long enough [19], in agreement with the observed excess of such systems over a wide range of periods [4, 27].

The properties of binaries formed by disk fragmentation depend on many parameters and the basic mechanisms are complex and sometimes not well understood [24, 28, 29]. These include the range of disk-to-star mass ratios and of the accretion of mass onto the disk from the parental cloud, the threshold for the onset of disk instabilities, the migration of secondaries within the disk, the accretion rates on the stars, the loss of angular momentum related to magneto-hydrodynamical winds, and the role of ternary or higher multiplicity systems. Given the large theoretical uncertainties, an accurate characterization of the binary population is a key requirement to constrain formation models.

Finally, it should be noticed here that the typical size of evolved (Meeus type II, [30]) proto-planetary disks (of the order of 300 au: [31–33]) is larger by an order of magnitude than the peak of the log-normal distribution of binary separation for Solar type stars (of the order of 40 au: [3, 4, 6]) and that we expect a continuum in the distribution of mass of companions formed by both cloud and disk fragmentation from stellar to substellar objects (see, e.g., [34–36]), a circumstance that is confirmed by observational data

Silvano Desidera, Francesco Marzari and Mariangela Bonavita contributed equally to this work.

[a] e-mail: raffaele.gratton@inaf.it (corresponding author)
[b] e-mail: silvano.desidera@inaf.it
[c] e-mail: francesco.marzari@unipd.it
[d] e-mail: mariangela.bonavita@open.ac.uk

Springer

[37–40]. Massive substellar companions may then likely form by these mechanisms [36, 41, 42] in the same separation range than planets but on a much shorter timescale than the usual core accretion mechanism invoked for planets [43–45], provided suitable conditions exist. Given the large incidence of binaries, we then expect that their formation plays a key role in planet formation and evolution [46–48], since it can both be considered as a competitive mechanism to core accretion—likely depending on the physical conditions in the disk early on in the stellar formation phase—and a precursor mechanism setting the stage for the late evolution of the disk when core accretion may still occur.

Enormous observational progress has been done in the last decade providing a wealth of new data about the formation of multiple stars. This includes among others the overwhelming size and accuracy of the samples allowed by the Gaia satellite, the contribution by ALMA in revealing cold circumstellar material, and last but not the least the advent of extremely performing high contrast imagers and NIR interferometers. In this review, we focus on the contribution given by one specific instrument, SPHERE at the Very Large Telescope [49], that is one of a battery of high contrast imagers built for the largest telescopes in the last decade that also includes GPI at Gemini [50], SCeXAO at Subaru [51], LMIRCAM at LBT [52], and MagAO at Magellan telescopes [53]. However, it should be clear that most important advances are done when combining information from the whole set of available instrumentation, so that speaking of results obtained with SPHERE also requires discussing the complementary constraints derived by the complete arsenal of such instruments (with JWST, Nancy Roman Telescope, and ELT contributions expected in the next few years).

In Sect. 2, we discuss the niche of high contrast imaging for observation of binary star. In Sect. 3, we summarize results from surveys of binary stars made using SPHERE. In Sects. 4 and 5, we present the contribution of SPHERE to the observation of stellar companions of disk hosting stars and to the presence of stellar companion to stars hosting planets/brown dwarfs. Finally, some conclusion is drawn in Sect. 6.

## 2 The high contrast imaging niche

Short period binaries are traditionally discovered as spectroscopic or eclipsing binaries and long period ones as visual binaries, with a large gap in between, partially filled by full-field interferometric observations, that are however time consuming and only sensitive to rather massive companions.[1] Progress in high precision radial velocities (RVs), in high contrast imaging, and in high precision space astrometry with Gaia is now filling this gap, providing a fairly comprehensive view of binary systems even at moderately large distances from the Sun. It is important to remind here that Gaia DR3 not only yielded more than 100,000 orbits for spectroscopic binaries and accelerations for about half a million astrometric binaries, but it is also providing very accurate maps of companions at very large separations, typically of the order of several hundreds to thousands au. These maps are complete down to $M \sim 0.08$ M$_\odot$, close to the Hydrogen burning limit, at 100 pc distance.

Figure 1 compares the regions around a star at 100 pc where various techniques typically used to detect companions are sensitive. The distance of 100 pc was considered here because it is the typical distance of the closest star forming regions. For comparison, we also show the period distribution of companions expected using the relation of [4]—assuming a solar mass and the relation between apparent separation and semimajor axis considered by [3]. At this distance, high contrast imaging explores the region between a few to some hundreds au, near the peak of the binary distribution, where the transition between binaries mainly formed by disk and cloud fragmentation occurs. Detections through high contrast imaging matches well that through Gaia astrometry (allowing, e.g., to solve the distance/mass degeneracy present for wide companions when using this technique) and fills the gap between discoveries by radial velocities and Gaia detections as visual binaries. Since high contrast imaging may well be complete down to the Hydrogen burning limit (and even further, in the substellar regime; see Fig. 2), we expect that it can detect a large fraction of the stellar companions; indeed, combining high contrast imaging with Gaia detections at large separation and spectroscopic binaries at short separation, virtually all stellar companions can be detected at this distance from the Sun.

An important caveat at very young ages is that a number of companions may be highly extinct or obscured by disc material (either circumprimary or circumsecondary or both). Examples include FW Tau C [55, 56], CS Cha B [57, 58], CrA-9 B [59]—with the latter two imaged with SPHERE. All these sources are significantly under-luminous (these examples present similar NIR photometry as expected for planets) but the shape of their spectrum (or ALMA kinematics for FW Tau C) suggest they are obscured M-dwarfs, with likely an edge-on disc blocking most of their light. These may trace the observable tip of obscured binary companions, but similar companions either closer from their star or more obscured would be missed with high-contrast imaging, hence may still be lurking undetected either in the large cavity of other transition discs (e.g., IRS 48). Similar objects at shorter separation may result in a potential lack of completeness at very young ages.

While the region around stars where they may be used is similar, high contrast imaging provides deeper contrasts than speckle interferometry, often used to study binary systems, at the price of a more complex (but still feasible) instrument that requires more calibrations—and then a bit slower in gathering data. When used at an 8m telescope, typical limiting contrasts for speckle interferometry are 6 to 8 mag in the optical bands at separations of a few tenths of arcsec (see [60]), while a high contrast imager

---

[1] When reasonably accurate positions are known, efficient fringe tracking may allow detections of very faint targets using interferometry, as obtained, e.g., with Gravity at VLTI [54].

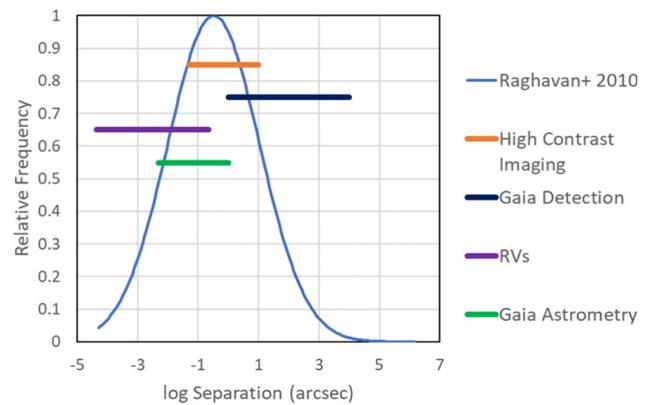

**Fig. 1** Distribution of binaries with separation at a distance of 100 pc (from [4]). Overimposed are segments representing schematically the range of separations at which various binary detection techniques (high contrast imaging, Gaia direct detection, radial velocities RVs, Gaia astrometry) are sensitive. High contrast imaging matches well Gaia astrometry and fills the gap between radial velocities and Gaia direct detection, near the peak of the binary distribution

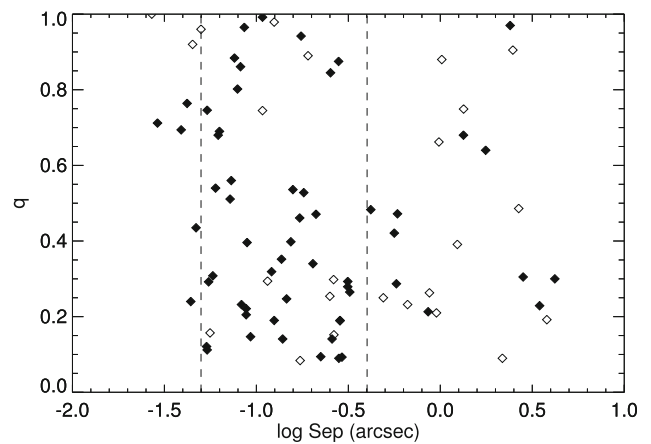

**Fig. 2** Distribution of the mass ratio $q$ in binary systems discovered within the high contrast imaging SHINE survey with separation (from [61]). Filled symbols are for new companions discovered in that paper; opens symbols are for previously known companions

may well go beyond 10 mag in the J and H bands at this separation even with very short exposures (see [61]). For a solar type star, this means that while speckle interferometry may detect companions of $> 0.16$ $M_\odot$, high contrast imaging is sensitive to objects less massive than the Hydrogen burning limit, down to planetary-mass objects for young stars, providing a complete picture of all stellar companions at such apparent separation. On the other hand, time series obtained putting together speckle interferometry and high contrast imaging data are very precious to follow the orbital motion.

High contrast imagers may also be used for determining orbits of binaries, with the final goal of determining accurate values for the orbital parameters, more specifically eccentricity and orbital inclination, and masses. Binaries generated by disk vs cloud fragmentation are expected to have systematically different parameter distribution. This does not only concern separation (and then period) and mass ratio, but moreover orbit eccentricity. In fact, binaries generated by cloud fragmentation are expected to have typically high eccentricities, close to a thermal distribution ($f(e) \sim 2e$: [62]), while lower eccentricities are expected for binaries generated by disk fragmentation (see discussion in [63]). We then expect a trend for increasing eccentricity with separation that is indeed confirmed by observations [64, 65]; these suggest that binaries with separations larger than 1000 au mostly formed by cloud fragmentation. Relative orbit inclination and predominance (or less) of corotation is also important for systems with higher multiplicity [66].

Accurate masses are especially important for low-mass very young stars, whose pre-main sequence evolution is poorly known. For instance, ages for young stellar groups determined using low-mass stars are typically younger than those obtained using more massive stars [67–70], a fact that is usually attributed to the presence of extensive active regions on their surfaces altering their colors [71, 72]. Since pre-main sequence stars are typically at quite large distances (tens of pc in the most favorable cases), the apparent separation of binaries with periods of a few years (and semimajor axis of a few au) are of the order of $\sim 0.1$ arcsec or even less. This is a regime where overlaps exist between high contrast imaging and indirect dynamical methods (radial velocities and Gaia astrometry), disclosing the possibility to solve the degeneracy between masses and separation at long periods, and with inclination in radial velocities. This greatly reduces the error bars in mass determinations (see, e.g., [73]). Companions detected with multiple techniques may then be used as benchmarks for stellar evolutionary models that are useful in particular for the pre-main sequence phase.

## 3 Results from surveys using SPHERE

SPHERE [49] is the high contrast imager for the ESO Very Large Telescope in operation since end 2014. SPHERE includes an extreme AO module (SAXO+), providing a high Strehl ratio of up to $\sim 0.9$ in the H-band, a suite of coronagraphs—the most used being apodized Lyot coronagraphs—, and three science instruments: the dual band NIR imager/polarimeter IRDIS (with a field of view of $11 \times 11$ arcsec), the NIR integral spectrograph IFS (field of view of $1.7 \times 1.7$ arcsec), and the visual imager/polarimeter ZIMPOL (field of view of $3.5 \times 3.5$ arcsec). All these instruments are diffraction limited and allow detections of very faint sources close to the targets down to contrasts of $\sim 10^{-6}$ at a few tenths of an arcsec. For what concern the present context, SPHERE allows to detect stellar companions down to a separation of $\sim 40$ mas in non-coronagraphic observations, and down to $\sim 90$ mas in deep coronagraphic ones. Furthermore, SPHERE provides relative astrometry accurate to some 1–1.5 mas (see also [74]). Due to the combination of competitive performances, very stable environment, versatility, relatively easy use and data reduction, and large user community, SPHERE is contributing a large fraction of the publications from high contrast imaging in the last few years. A query to NASA-ADS done on June 28, 2022, yielded 275 refereed papers published after 2014 using the keywords "SPHERE" AND "VLT" in the Abstract,[2] with respect to 62 using "GPI" AND "Gemini", 44 using "SCEXAO" AND "Subaru", 25 using "LMirCam" and 30 using "MagAO".

As mentioned above, SPHERE is able to detect all stellar companions in its field of view with separation larger than 90 mas, and most of them for separation larger than 40 mas, at least if we exclude very young objects still embedded in the circumstellar disks, that will be discussed in the next Section. So, all observations done with SPHERE provide conclusive information about the presence or lack of stellar companions in this range of separation; for typical SPHERE targets this corresponds to the peak of the period distribution. However, programs with SPHERE were mainly focused on substellar objects and disks, and most of them excluded by design targets known to have stellar companions from previous studies; for instance, this was the case of the extensive SHINE survey [40, 75, 76] run by the SPHERE consortium within the guaranteed time observations. In spite of this, the unprecedented sensitivity of SPHERE lead to the discovery of 78 binaries (56 of which are new discoveries) in the sample of 463 stars observed within SHINE [61]. An exam of the properties of the input catalogue used by SHINE showed that considering also objects excluded a priori because known binaries, it was possible to set up a sample of young (age $< 700$ Myr) binaries for which the search of stellar companion is essentially complete for separation in the range 0.05–0.5 arcsec (roughly 5–40 au). The fraction of stars with binary companions in this range agrees very well with previous estimates for older stars made using spectroscopic binaries and visual binaries close to the Sun [3, 4]. The period distribution of SHINE binaries is clearly bimodal, with a main peak at low mass ratio and a second one for equal mass systems. In particular, SHINE was very efficient in discovering low mass ratio binaries, reducing the uncertainties in the completeness correction that affect previous determinations (see, e.g., [4]). Combining SPHERE observations with other determinations and the Hipparcos-Gaia Proper Motion Anomaly (PMA, see below) [77], preliminary orbital parameters were obtained for a significant fraction of these binaries; a moderate eccentricity seems favored in most cases, as expected for formation by disk fragmentation. However, while contributing to this area, SPHERE data alone are not however enough for a statistically significant result mainly due to the small fraction of the orbit followed. For the few objects for which such estimate was possible, the values of the masses derived from dynamical arguments were in good agreement with the model predictions. Stellar and orbital spins appear fairly well aligned for the 12 stars having enough data. All this data favors a disk fragmentation origin for most of the binaries in this separation range.

On the other hand, there is a substantial overlap of SHINE binaries with stars having large PMA obtained comparing Gaia proper motions with the long term proper motions obtained combining Hipparcos and Gaia positions [77–79]: 75% of the SHINE binaries that are also in Hipparcos are identified as binaries using PMA; and companions were found by SPHERE for 70% of the binaries from PMA that were observed. The remaining 30% objects with significant PMA and no detection in SHINE likely have companions at very small separation and/or very low mass ratio. High contrast imaging may break the distance-mass degeneracy inherent to the PMA detections. In addition, the large overlap suggests that selecting targets with large PMA may greatly enhance the detection yield of high contrast imaging surveys, that is typically low. This was exploited by the COPAINS program [80] that focused on young stars with the goal of detecting substellar objects. The original sample of 25 stars was actually selected before availability of Gaia DR2 data and used lower precision Tycho and ground-based proper motions for target selection. However, eight of these targets have significant PMA using Gaia eDR3 data [78]; the COPAINS program found the companion responsible for the observed astrometric acceleration in seven cases, four of them being Brown Dwarfs or stars close to the Hydrogen burning limit. The undetected object is likely a planetary mass companion at relatively small separation, below the detection limit of the survey. In addition, [80] noticed that once the time elapsed between the Gaia and SPHERE observing epochs is taken into account, the position angle of the detected object is compatible within the errors with the direction of the PMA vector, as it should be expected given that this is substantially a measure of the instantaneous acceleration of the star due to the presence of the companion.[3]

---

[2] This total is very similar to the value of 271 considered by the ESO bibliographic system up to end 2021 (see https://www.eso.org/sci/libraries/edocs/ESO/ESOstats.pdf).

[3] This is exactly true if the Hipparcos-Gaia proper motion really represents the long term proper motion of the star, a fact that is true only if the orbital period is shorter than or comparable to the time elapsed between Hipparcos and Gaia epochs, that is about 24 year.

For what concerns orbit and mass determination, [81] considered a sample of previously known 15 young low-mass binaries and combined high contrast data obtained with SPHERE with those at shortest wavelengths provided by AstraLux to obtain their orbits, and then masses of the components. Quite accurate results were obtained for seven systems, confirming that magnetic models better matches the observed colors. Additional interesting cases have been presented for other young low-mass systems [73, 82].

High contrast imaging is also particularly well suited to study systems of intermediate separation (a few tens au) formed by a main sequence star and a white dwarf. Several such systems, usually called Sirius-like and related to Ba- and CH-stars, [83, 84], have been studied with SPHERE [85, 86] and other instruments [87–89]. Some of the Sirius-like systems were spuriously considered as young because there is angular momentum transfer associated to the (wind driven) mass transfer occurring when the progenitor of the white dwarf is on the asymptotic giant branch, making the secondary a faster rotator than expected for its age. Combining SPHERE and high spectral resolution observations allows a fine discussion of the white dwarf progenitor and of the nucleosynthesis occurring within it [86, 90].

## 4 Stellar companions of disk hosting stars

Observations of stellar companions around stars still surrounded by the primordial, gas rich disk, provide snapshots of the binary formation phase and it is then crucial in understanding how binaries form. ALMA has been very successful in detecting such disks, far extending the capabilities of earlier sub-mm or radio instruments such as Plateau du Bure or VLA. Coupling these disk detections with observations of faint companions from, e.g., high contrast imaging or interferometry allows to study several interesting cases. Early systematic studies of the relation between multiplicity and proto-planetary disks combining ALMA and high contrast imaging data were performed by [93, 94]. Furthermore, high contrast imagers such as SPHERE [49], GPI [50] and SceXAO [51] offer the opportunity to observe both very faint close stellar objects exploiting Angular Differential Imaging related techniques, as well as disks and other structures seen in scattered light, mainly (but not exclusively) in polarimetry. Such observations have revolutionized this field in the last few years.

### 4.1 Outer stellar companions

The presence of nearby stellar companion(s) external to the disk (and not necessarily formed within the disk) can significantly affect the morphology of protoplanetary disk around very young stars; typically, lower fluxes are observed in polarimetric observations for the quite frequent cases of disk hosting stars that have a companion within a few hundreds au [95]. During the pericenter passage of the secondary star of eccentric binaries, strong spiral arms develop in the disk of both the primary and secondary star. This has been shown by hydrodynamical numerical simulations of close binary stars [91, 96–101] taking as test bench the $\gamma$ Cep and $\alpha$ Cen systems. In Fig. 3 the formation of the spiral arms is illustrated in 3D by the simulations performed by [91]. The spirals develop in the disk plane but can also be seen in the vertical direction (Fig. 3, bottom plot) and might be visible in edge-on disks. The observations of spirals in a disk may be a strong indication that the companion star is currently close to the pericenter of its orbit. Possible such cases are those of HD 100453 ([92]; see Fig. 4) and UX Tau [102] even if alternative explanations have been proposed for the features observed in these disks [103].

### 4.2 Embedded companions

Observations of stellar companions embedded within protoplanetary disks are of crucial importance to establish if they can really form by disk fragmentation and with which parameters. A few examples of such objects were observed with SPHERE (and other high contrast imagers). Since massive objects cause large dynamical perturbations that de-stabilize the disk on a quite short timescale, such observations are expected to be relatively rare and biased toward small secondary masses unless the separation between the components is very small. In a few cases, at the centre of SPHERE targets there are indeed very compact binaries. A typical example is V4046 Sgr that is an eclipsing binary T Tau variable made of two similar components [104]. In the near IR, the disk scatters the light from the central system and due to the mutual eclipses between the components, shadows are casted on the disk that rotate with the period of the central star [105]. This requires a good alignment between the orbit of the central system and the disk, that strongly argues for a common origin between the binary and the disk. A similar case (but for the lack of mutual eclipses) are the Herbig AeBe stars HD 34700 and AK Sco, that are equal mass short period spectroscopic binaries [106, 107], surrounded by a wide cavity and very bright disk with multiple spiral arms [108–110]. HD 34700 has two further out T Tau companions at several thousands au [111]. The close alignment between central close binaries and the circumbinary disk seems indeed to be a very frequent configuration [112, 113].

In other cases, the separation of the central binary is large. In general, in these cases the presence of the stellar companion causes a large cavity in the disk because of the strong dynamical perturbations (see, e.g., [118] and discussion therein). These objects have then a spectral energy distribution (SED) characterized by two peaks in the near- and far-IR, one due to the photosphere and the circumstellar disk(s) and a second one by the circumbinary disk, mimicking the case of transition disks [119, 120]. A famous case is the F-type star HD 142527. The SPHERE observations were discussed in [121–123], who determined that the small mass companion previously found by [124] is on a highly eccentric orbit, that justifies the wide observed cavity. This disk is very complex, with the presence of shadows and horseshoes that all can be justified by the binarity of the star [125]. A similar case is that of the Herbig Be star MWC 297 [126], where the SPHERE observation shows a small mass (0.1–0.5 $M_\odot$) companion within the gap in the disk whose presence is shown indirectly from the SED. Similar cases not observed by SPHERE includes IRAS 04158+2805 [127]. Incidentally, we notice that while cavities and other features (horseshoes, spirals) are generally present in these disks, there are several systems with disks presenting such features but where the companion has not been actually detected (see, e.g., the case of IRS 48 [128]).

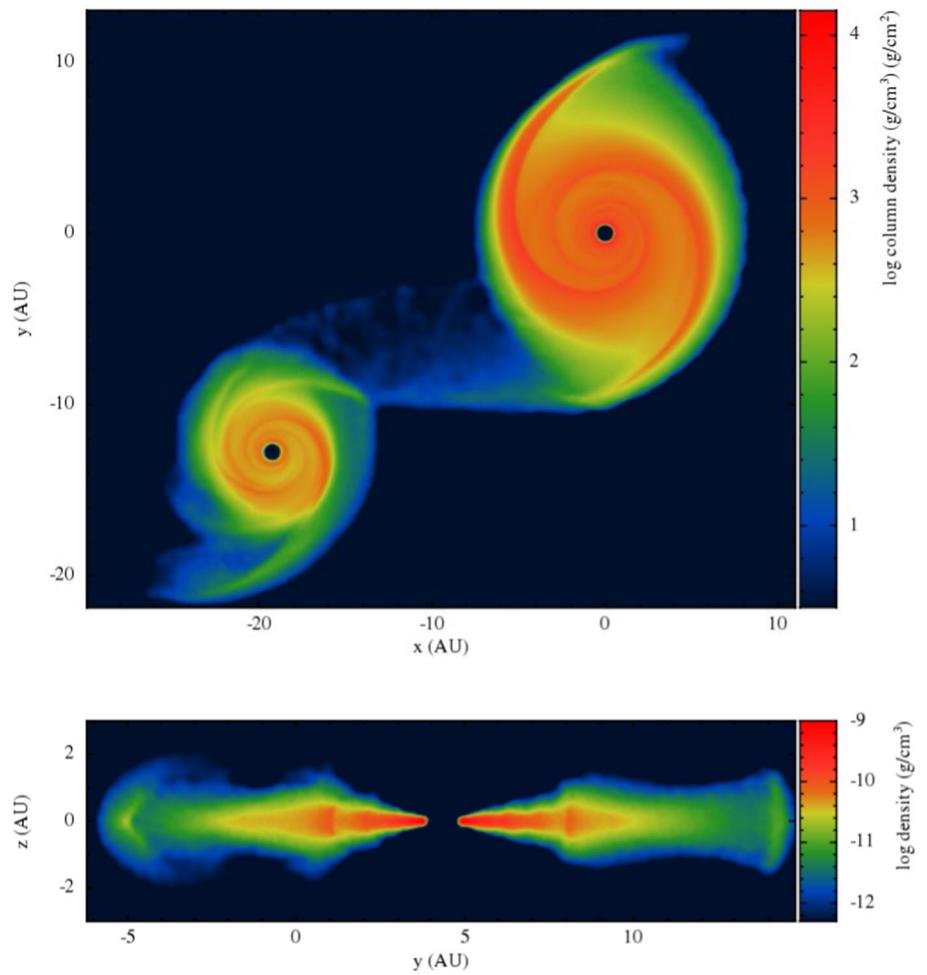

**Fig. 3** 3D modeling of the circumstellar disks surrounding the components of a close binary system [91]. In the top panel the column integrated superficial density is shown during the pericenter close passage. On the bottom panel and *x*–*z* section of the circumprimary disk is shown at the same time

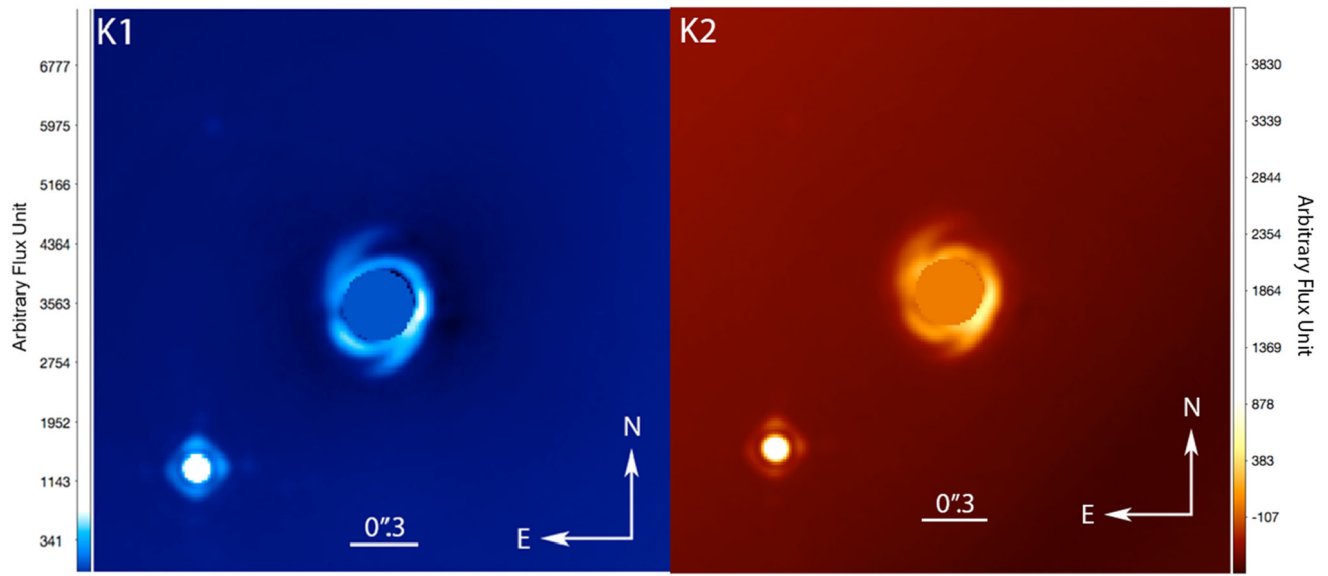

**Fig. 4** SPHERE observation of HD100453 [92]. Note the spiral arm design and the faint stellar companion SE of the star

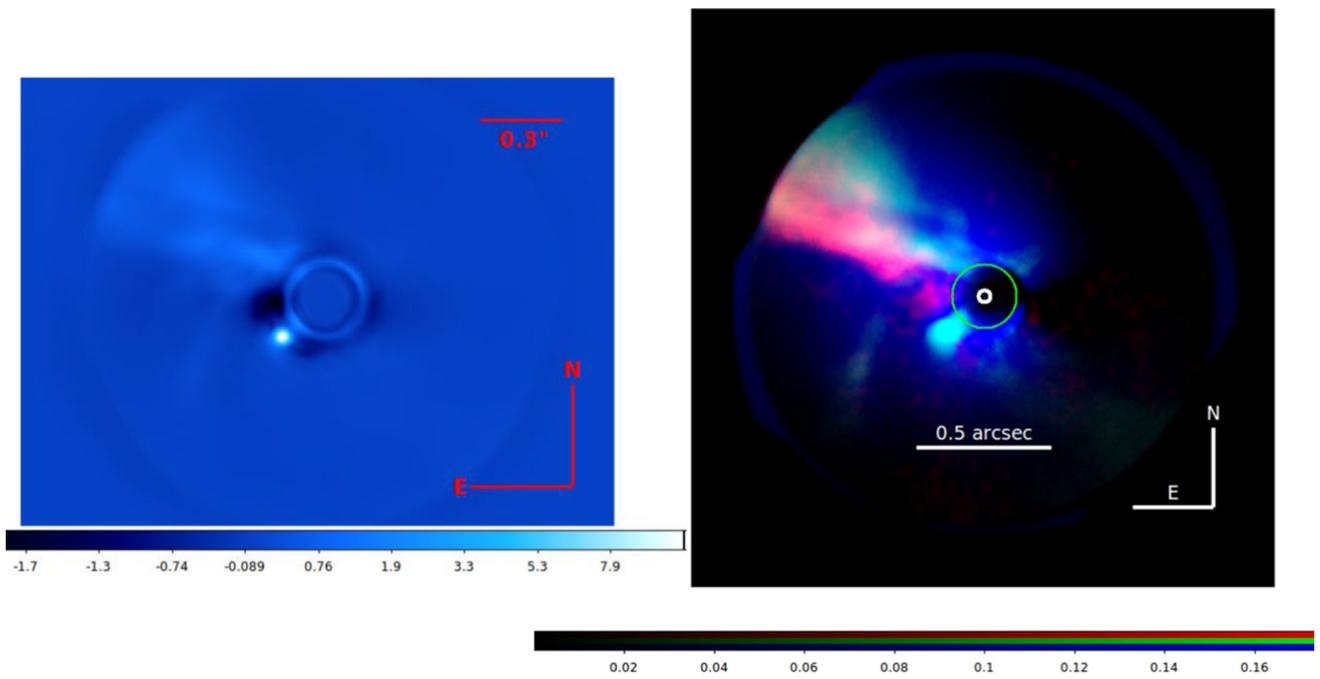

**Fig. 5** Images of R CrA obtained combining observations obtained with SPHERE IFS in the NIR. N is up and E is to the left. Different scales have been used in the two images to show different features. Left: image from [114] showing only very illuminated pixels, clearly showing the small mass stellar companion about 0.2 arcsec SE of the central (unresolved) binary, here hidden by the coronagraphic mask. . Right: combined color images from [115] showing the weak signal at different wavelengths. The prominent feature NE of the star is the outflow. The stellar companion image is saturated here

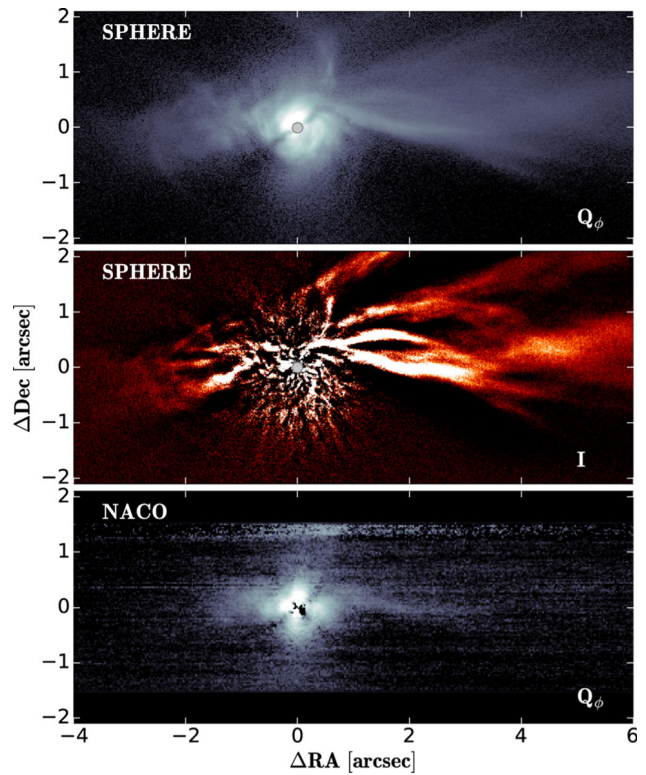

**Fig. 6** SPHERE and NACO polarized images of SU Aur. North is up and east is to the left. From [116]

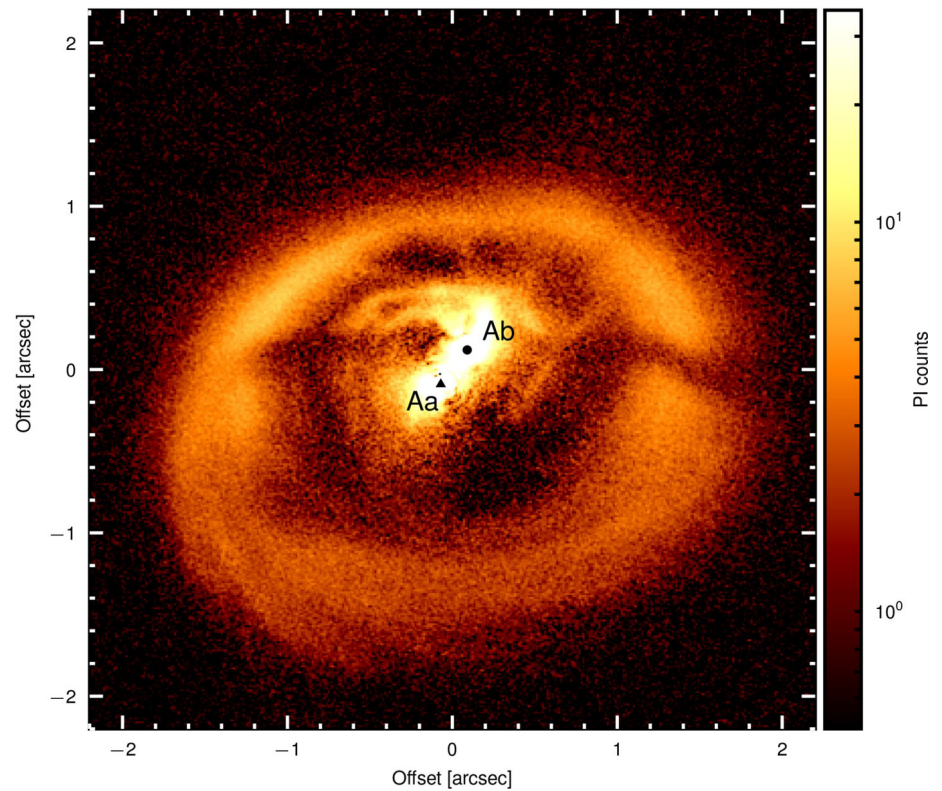

**Fig. 7** SPHERE polarized intensity image of GG Tau A. The image is centered on the expected location of the system's center of mass. The locations of GG Tau Aa and Ab are marked by a black triangle and circle, respectively. North is up and east is to the left. From [117]

Disk hosting stars with even higher multiplicity are also observed. An interesting example is R CrA (see Fig. 5) that is composed of an intermediate mass central binary with a period of 56 d and semimajor axis of $\sim 0.4$ au [129] around which there is a circumbinary disk. The system is seen almost edge-on, so that the central binary is partially occulted by the disk; due to the binary motion, the system is variable with the orbital period because alternatively either of the two components (that have similar but not identical mass) emerges from the disk and it is then visible. A third low mass component is also present further out than the circumbinary disk [114] at a few tens of au from the central binary, and there is hint for a disk segment even further out. A notable outflow is also seen in scattered light [115], roughly perpendicular to the disk; all system components looks then quite well aligned. An even more complex pattern is shown by T Tau, that is a triple likely hierarchical system for which spectacular images have been obtained using SPHERE [130]. While there is a lot of circumstellar material around this system, the geometry is very complex and not well understood, possibly suggesting the superposition of several different components (circumbinary disk and outflows from the various components).

While the majority of systems appear to be aligned, there are cases where this is not the case. Spectacular cases are those of SU Aur ([116], see Fig. 6), and GG Tau ([117]; see Fig. 7) where the disks show intricate spiral structures and shadow lanes, cast by inner, misaligned disk components.[4] A similar case, but seen on a different orientation, is T CrA (Rigliaco et al., in preparation). In this last case a circumbinary disk is observed; both the binary orbit and the disk are seen at high inclination, but at very different position angles—indicating a large mutual inclination. As in the case of R CrA, the stellar binarity causes periodic variations in the source, but in this case the period is much longer (29.6 year), so that a disk around the primary is likely present, as shown by the SED. A very prominent outflow is observed related to molecular hydrogen emission-line objects [132], that are present because the star is embedded in a dense interstellar environment. The outflow is roughly perpendicular to the binary orbit but not to the circumbinary disk. SU Aur, GG Tau and T CrA may be likely explained by late accretion episodes, fuelled by material having a very different angular momentum with respect to that responsible for the accretion of most of the mass on the system. Still, at least in the case of T CrA the star is projected close to the filament seen in Herschel data [133], suggesting that the large change in the direction of the angular momentum vector is actually due to a misalignment of the position of the star with respect to the filament (of the order of 1000 au or less) that it is actually small if compared to the filament width that is at least an order of magnitude larger (see, e.g., [134]). This shows the relevance of the huge scale range involved in star formation.

---

[4] An additional case of misalignment between a circumbinary disk and an inner wide binary is that of HD 98800 [131]; however the disk around this system is a debris rather than a gas-rich disk.

## 5 Stellar companions of planet/brown dwarf hosting stars

5.1 Wide stellar binaries

Planet formation in binary star systems is strongly influenced by the gravity of the companion star. The effects range from increasing the temperature of the disk via shock waves at the spiral arms [91, 96] making it difficult to condense ices even beyond the frost line, to a vertical remixing of the dust grains due to hydraulic jumps developing at the location of the spiral arms [91] and to the excitation of the eccentricities of the planetesimals during their accumulation [135–137] only partly helped by the pericenter alignment. In a recent paper [138] show via numerical simulations that planetesimal growth may be promoted in many parts of the disk in a binary. However, this occurs only if initial planetesimal seeds as large as 1–10 km are already present in the disk, a condition requiring additional ways to produce planetesimals such as instabilities in the disk that might be favored by the companion perturbations. However, modeling the formation and evolution of planets in binaries appears to be a challenging task due to the wide parameter space to explore like the binary mass ratio and the semi-major axis, eccentricity and inclination of the binary orbit.

SPHERE and high contrast imaging instruments are contributing to this topic by dedicated searches for companions, mostly stellar, targeting stars with planets discovered by other techniques, such as radial velocities (RV) and transits. These stars are typically old, reducing the sensitivity of imaging observations for substellar objects. Moutou et al. [139] observed 68 stars with RV planets, finding a larger frequency of stellar companions for stars with planets in highly eccentric orbits and confirming the paucity of close binaries (separation < 50–100 au) among stars with RV planets. Ginski et al. [140] detected with SPHERE four new stellar companions, including a white dwarf, at close separations from a survey of 122 RV planet hosts. A by-product of imaging searches with high sensitivity at very close separations is represented by the cleaning of the lists of substellar objects identified with the RV technique from false alarms due to stellar companions with minimum mass $m \sin i$ in the substellar regime because of pole-on orbits. One such case is represented by HD 211847, whose RV companion with $m \sin i = 19\,M_J$ is likely the same object seen by SPHERE at 0.22" = 11.3 au with a photometric mass of 0.15 $M_\odot$ [139]. Cheetham et al. [141] also identified a T-type brown dwarf (with some tension between mass derived from dynamical methods and luminosity) around the RV planet host HD 4113.

Host stars with transiting planets have also been observed with SPHERE. Bohn et al. [142] identified several candidates, with at least nine binary companions in the field of view of the SPHERE instrument in their sample of 45 stars with Hot Jupiters. Desidera et al., A &A, submitted, identified with SPHERE observations a close companion at stellar/substellar boundary, with mass from luminosity $83^{+4}_{-6}\,M_J$ at a projected separation of 84.5 mas =3.3 au around TOI-179, a young K dwarf hosting a compact Neptune-mass transiting planet in close and eccentric orbit. Unfortunately, even if coupled with the dynamical signatures (RV long term trend and proper motion anomaly), the single-epoch imaging observation does not allow a full orbit determination; it is possible that significant dynamical interactions are or were at work in this system. The complex architecture of the system is completed by the presence of a close pair of K dwarfs at a projected separation of 3400 au.

5.2 Circumbinary planets

Less difficult appears to be planet formation in circumbinary configurations. Due to the binary perturbations, planetesimal accumulation close to the centre of the system is strongly inhibited but beyond a given distance from the star planets can assemble from the dust [143]. These planets may later migrate inwards by interaction with the gaseous disk and end up in orbits close to the central binary, often close to mean motion resonances [144, 145]. Taking into account the complexities of forming planets in binary star systems and their peculiar dynamical evolution it is important to find potential stellar companions to planet hosting stars.

SPHERE contribute to this area through deep search for planets in wide orbits around stars which are known to be binary systems, either before the imaging observations or as results of further follow-up after planet discoveries. Dedicated searches have been performed around very close binaries, typically of young ages, looking for circumbinary planets [146] and around individual members of moderately wide binaries [147]. When considering the multiplicity characteristics of stars with planets and low-mass BDs detected in direct imaging (DI), it should be noticed that the age distribution is biased toward small values (because of the better sensitivity as planets are brighter at young ages). Several of these objects are members of young moving groups or association. Therefore, there is some ambiguity among comoving objects and bound companions at extremely wide separations [148, 149]. Nevertheless, the observational results indicate that outer stellar companions are rare. Among robust cases of outer stellar companions we mention 51 Eri, which forms a triple system with the close binary GJ 3305 [150, 151]. HD 74865 hosts a wide, low-mass stellar companion [149] beside the close directly-imaged low-mass BD [152]. In the case of ROX 42B, the primary ROX-42 (at a projected separation 79" ∼11500 au from ROX-42B), is a tight triple system [the star is an SB2 with another visual component at 0.4", 153, 154]. Other cases of additional very wide stellar companions around BDs companions detected with DI are represented by GQ Lup [155] and DH Tau [156].

The low frequency of wide stellar companions around stars with DI planets is probably linked to the fact that the period range corresponding to the peak in the distribution of binary companions is already occupied by the planet themselves. We also stress that availability of deep AO imaging (up to the field of view of the instruments) and the good sensitivity of Gaia and other all-sky surveys at larger separation (thanks to the young age) ensure a good detection completeness.

The situation is different for what concerns close stellar companions within the planet or BD orbits. Several cases of planets and BD in circumbinary configuration are known [157], e.g., TWA 5 [158], HD 106906 [157, 159], HIP 79098 [160], and b Cen [161]. In some cases the inner binary orbit is not properly characterized. On the other hand, it should be considered that in some cases RV monitoring reveals very close companion around DI planet hosts is missing or have poor sensitivity because of fast rotation and early spectral type of many of them [see, e.g., 162]. Therefore, some observational incompleteness is expected. The current evidences is that the frequency of substellar companions in circumbinary configuration is consistent with that of the same objects around isolated host stars [146].

A special case, in some way merging the two science cases above, is represented by HD 284149 [163]. The substellar companion at wide separation (3.67") was originally discovered by [164]. Follow-up observations with SPHERE, mostly aimed at the characterization of the brown dwarf, identified a low-mass, stellar companion (0.16 $M_\odot$) at 0.1" from the star.

## 6 Conclusions

Our view of the formation of the ubiquitous binary stars is evolving due to the contribution of new highly performing instruments. In addition to Gaia and ALMA, a special role is that of high contrast imaging such as SPHERE. These instruments allow a close view of the region around the peak of the distribution with separation of binaries (tens to hundreds au) at the distances of the closest star forming regions. This is important because with these instruments we may directly see very young or even still embedded binaries during their formation phase. A picture is emerging where disk fragmentation and instability plays a crucial role and there may be a continuum between the formation of substellar objects (massive planets and brown dwarfs) and of stellar companions, up to equal mass objects. Observations indicate that the mechanisms involved are complex; they include accretion from interstellar filaments onto the disk, the generation of instabilities in the disks, accretion on the components of multiple systems, migration due to interactions with the disk, dynamical interactions between multiple systems. The interplay between all these mechanisms generate the variety of observed systems. Various unexpected features are emerging, such as the relatively high frequency of misaligned systems. While up to now the number of systems studied in detail is still rather limited, accumulation of new data is progressing and in the near future we may expect the first statistical studies trying to capture the huge variety of the observed systems in a better understanding of the relative roles of the most relevant mechanisms.

**Acknowledgements** This work has made use of the SPHERE Data Centre, jointly operated by OSUG/IPAG (Grenoble), PYTHEAS/LAM/CeSAM (Marseille), OCA/Lagrange (Nice), Observatoire de Paris/LESIA (Paris), and Observatoire de Lyon/CRAL, and supported by a grant from Labex OSUG2020 (Investissements d'avenir—ANR10 LABX56). This work has been supported by the PRIN-INAF 2019 "Planetary systems at young ages (PLATEA)" and ASI-INAF agreement n.2018-16-HH.0.

**Funding** Open access funding provided by Istituto Nazionale di Astrofisica within the CRUI-CARE Agreement.

**Data Availability** Data sharing not applicable to this article as no datasets were generated or analyzed during the current study.